\documentstyle[sprocl,epsfig]{article}

\def\Journal#1#2#3#4{{#1} {\bf #2}, #3 (#4)}

\def\NPB{{\em Nucl. Phys.} B}

\def\PRD{{\em Phys. Rev.} D}

\def\EPJC{{\em Eur. Phys. J.} C}

\def\be{\begin{equation}}
\def\ee{\end{equation}}
\def\bea{\begin{eqnarray}}
\def\eea{\end{eqnarray}}

\newcommand{\vg}{\gamma(P^2)}

\begin{document}
\title{QCD ANALYSIS OF STRUCTURE FUNCTIONS OF REAL AND VIRTUAL 
 PHOTONS\footnote{Talk given at the DIS-2000 conference, Liverpool, 
 25-30 April 2000}}

\author{I. Schienbein}

\address{Institut f\"ur Physik, Universit\"at 
         Dortmund\\ 
         D--44221 Dortmund, Germany\\ 
         E-mail: schien@hal1.physik.uni-dortmund.de}
\maketitle\abstracts{ Parameter--free and perturbatively stable 
 leading order (LO) and next--to--leading order (NLO) parton densities 
 for real and virtual photons are presented.}
\section{Introduction}
The hadronic structure of real and virtual photons can be measured in 
electron positron scattering processes in which one lepton serves as a source of
target photons with virtuality $P^2 << Q^2$ where $Q^2$
is the virtuality of the probing photon.
The measured $e^+ e^-$ cross section can then be obtained by a convolution
of a flux of target photons with the deep inelastic electron (positron) 
photon scattering cross section which is (completely analogous to the 
case of deep inelastic electron proton scattering) described by two 
structure functions $F_{2,L}$.
In the QCD impproved parton model the photon structure functions 
$F_{2,L}^{\vg}(x,Q^2)$ are given by a convolution of non-perturbative 
parton densities of the (real or virtual)
photon with appropriate perturbatively calculable short distance 
coefficient functions.
The photonic parton distributions are governed by inhomogenous evolution
equations which have to be supplemented with appropriate boundary conditions.

Here we briefly describe recently proposed parameter--free leading order (LO) 
and next--to--leading order (NLO) 
boundary conditions for real and virtual photons \cite{GRSvg99}.
\section{Boundary Conditions}
The boundary conditions are given in the DIS$_\gamma$ scheme 
at a low resolution scale $Q_0^2\approx 0.3$ GeV$^2$. 
The exact LO and NLO values of the universal (i.\ e.\ hadron--independent) 
input scale $Q_0$ are fixed by the experimentally well constrained radiative 
parton densities of the proton \cite{GRV98}.
\subsection{Real Photon}
The boundary conditions for the real photon \cite{GRSvg99} are given
by a vector meson dominance (VMD) ansatz where (at the low scale $Q_0$)
the physical photon
is assumed to be a coherent superposition of vector mesons which
have the same quantum numbers as the photon
\begin{equation}
f^{\gamma}(x,Q_0^2) = f_{had}^{\gamma}(x,Q_0^2) = 
\alpha G_f^2 f^{\pi^0}(x,Q_0^2)
\label{input}
\end{equation}
with $G_{u,d}^2 = (g_{\rho} \pm g_{\omega})^2$ and $G_g^2 = G_s^2 
= g_{\rho}^2 + g_{\omega}^2$ ($g_{\rho}^2=0.50$, $g_{\omega}^2=0.043$). 
This optimal coherence maximally enhances the up quark which is favoured by the 
experimental data.

Since parton distributions of vector mesons ($\rho, \omega$,...) are 
experimentally undetermined, we furthermore assume that these are 
similar to pionic parton distributions.
Thus, for given pionic parton distributions our model has no free parameter.
\subsection{Pion}
Since only the pionic valence density is experimentally
rather well known, we utilize a constituent quark model \cite{Alt74} to relate
the pionic light sea and gluon to the much better known parton distributions
of the proton \cite{GRV98}.
In Mellin-$n$-space one easily finds the following boundary conditions for 
the pion distribution functions \cite{GRS98,GRSpi99}
\begin{equation}
g^{\pi}(n,Q_0^2) = \frac{v^{\pi}}{v^p}\, g^p,
\quad\quad
\bar{q}\,^{\pi}(n,Q_0^2) = \frac{v^{\pi}}{v^p}\,
\bar{q}\,^p,
\end{equation}
which only depend on the rather well determined valence distribution 
of the pion and the parton distributions of the proton.
\subsection{Virtual Photon}
In the virtual photon case we propose the following boundary conditions
which of course smoothly extrapolate to the real photon case  
($P^2 \rightarrow 0$) \cite{GRSvg99}:
\begin{equation}
f^{\gamma(P^2)}(x,Q^2=\tilde{P}^2) = 
f_{\rm{had}}^{\gamma(P^2)}(x,\tilde{P}^2) = \eta(P^2) 
f_{\rm{had}}^{\gamma}(x,\tilde{P}^2) \, .
\end{equation}
The employed dipole suppression factor (rho-meson propagator) 
$\eta(P^2)=(1+P^2/m_\rho^2)^{-2}$ is somewhat speculative and can
be regarded as the simplest choice of modelling the $P^2$--suppression.
\section{Numerical results}
\begin{centering}
\begin{figure}[thb]
\epsfig{figure=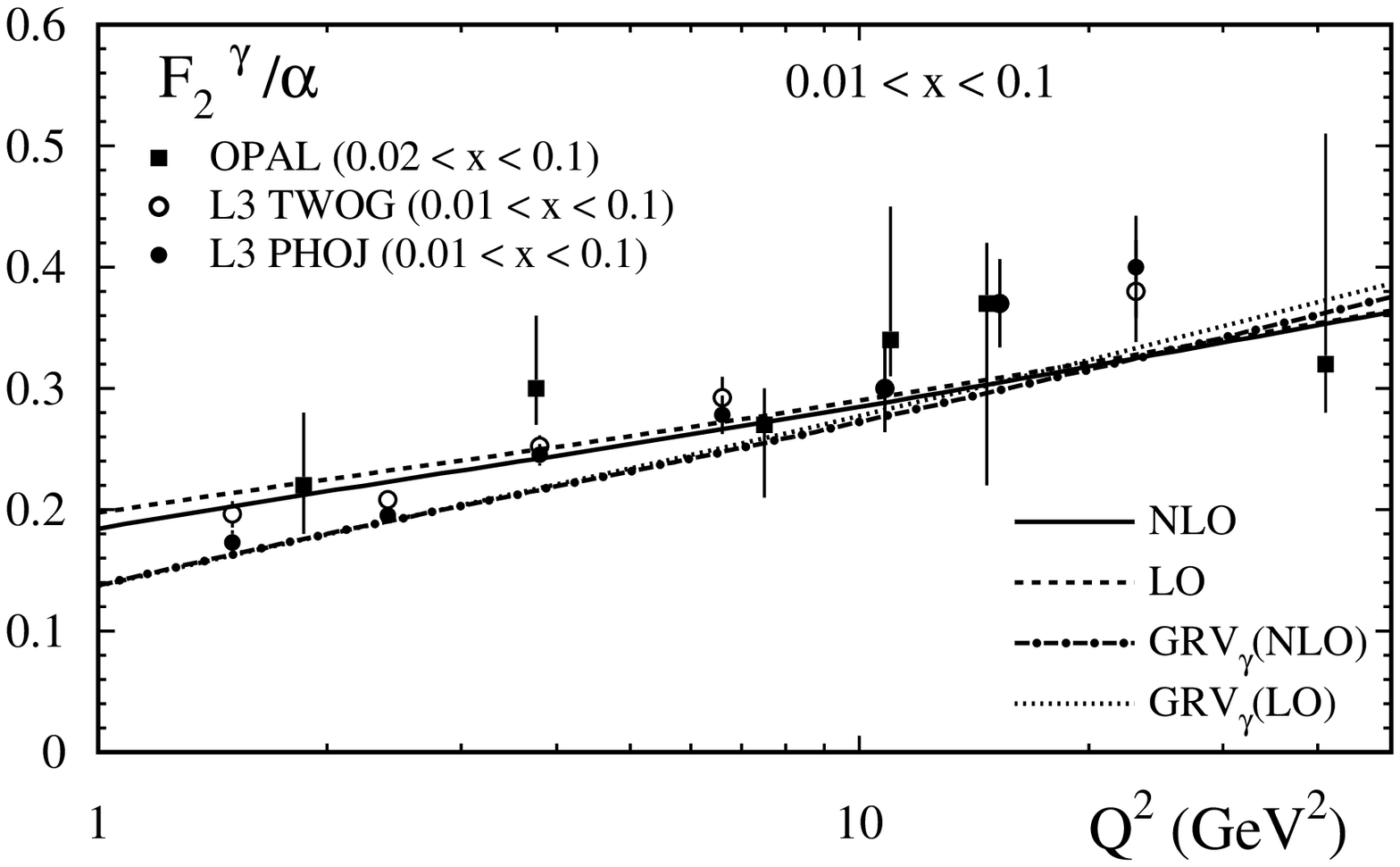,width=5.5cm}
\epsfig{figure=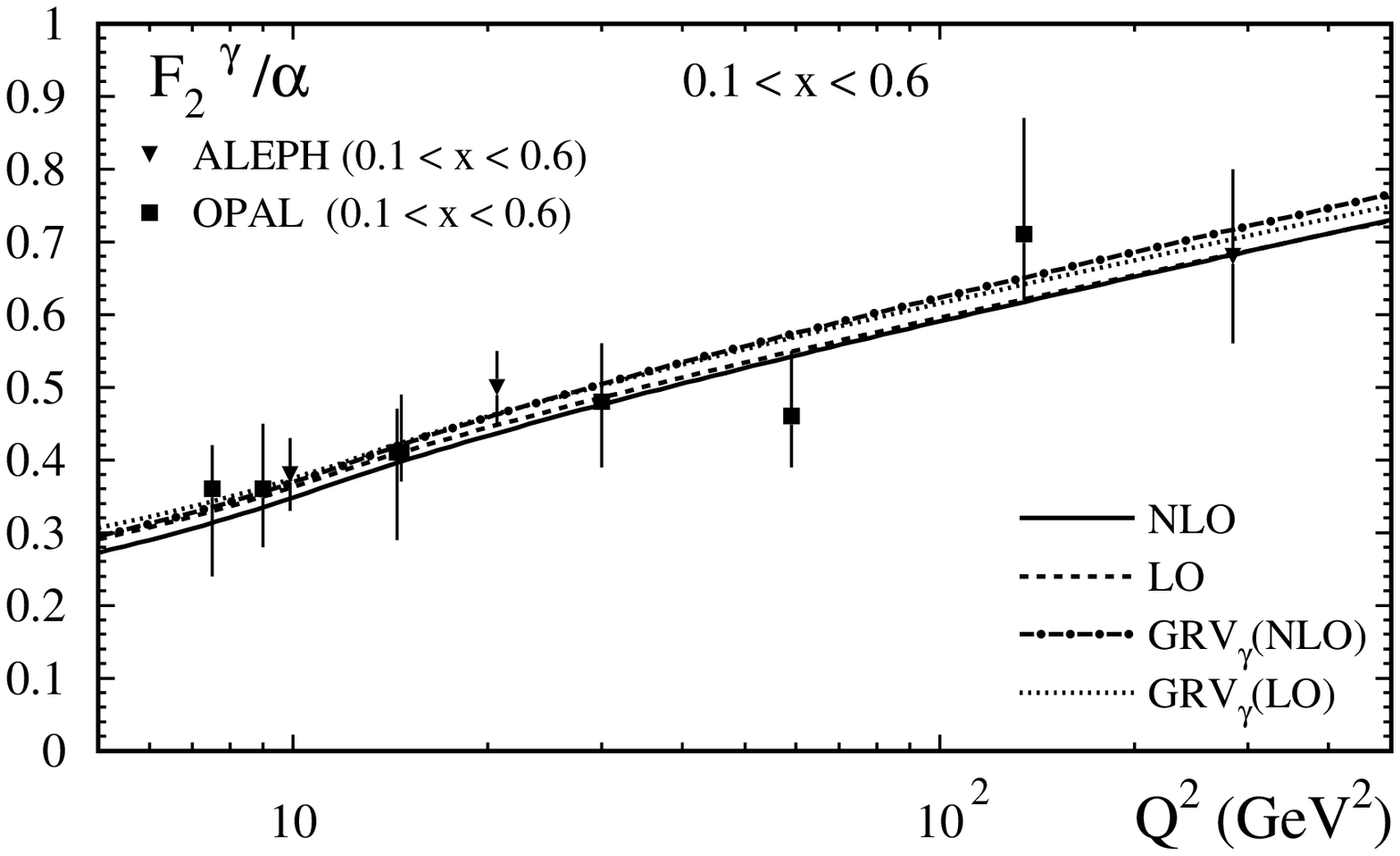,width=5.5cm}

\vspace*{-0.5cm}
\caption{\small $F_2^{\gamma}/{\alpha}$ in dependence of $Q^2$ for two
different $x$-bins.} 
\label{fig:q2evol}
\end{figure}
\end{centering}
The presented model successfully describes LEP data on the photon structure
function $F_2^\gamma$ and H1 dijet data. 
For details see Ref. \cite{GRSvg99} and references therein.
As an example, in Fig. \ref{fig:q2evol} the evolution
of $F_2$ with $Q^2$ is shown for two different $x$-bins.
Also shown are the GRV$_\gamma$ predictions \cite{GRVg92}. 
For $0.1 < x < 0.6$ all curves are close together because this $x$-range
is already dominated by the point-like solution, whereas for  $0.01 < x < 0.1$
our results are larger at small values of $Q^2$  and evolve weaklier with $Q^2$ due
to the different boundary conditions.
\section*{Acknowledgments}
Supported in part by the {\it Gra\-du\-ierten\-kolleg 'Erzeugung und Zer\-f\"alle 
von Ele\-mentar\-teil\-chen'} of the {\it Deutsche Forschungs\-gemein\-schaft} 
at the {\it Uni\-versit\"at Dortmund} and by the {\it Bundes\-ministerium f\"ur
Bildung, Wissenschaft, Forschung und Technologie}, Bonn.

\end{document}